\documentstyle[preprint,aps,prb]{revtex}
\begin{document}
\baselineskip 0.7cm
\title{Excitation spectrum
of the  S=1/2 quantum spin ladder with frustration: elementary
quasiparticles and 
many-particle bound states.
}
\author{V.N. Kotov$^{1}$, O.P. Sushkov$^{1}$, and R. Eder$^{2}$}
\address{$^1$School of Physics, University of New South Wales, Sydney 2052,
 Australia\\ 
$^2$Institut f\"ur Theoretische Physik, Universit\"at W\"urzburg,
 Am Hubland,  97074 W\"urzburg, Germany
}

\date{\today}
\maketitle
\begin{abstract}
\baselineskip 0.6cm

The excitation spectrum of the  two-chain  $S=1/2$ Heisenberg spin ladder
 with additional second neighbor
 frustrating interactions  is studied by a variety of
techniques. A description, based on a mapping of the model onto a
 Bose gas of hard-core triplets is used to determine the one- and
two-particle excitation spectra. We find that low-lying singlet and triplet
bound states are present and their binding energy increases with
increasing frustration. 
In addition, many-particle bound states are found by 
exact diagonalization and variational methods. We prove that
the larger the number of bound particles the larger the binding
energy.  Thus the excitation spectrum has a complex structure 
 and  consists of elementary triplets and composite many-particle
 singlet and triplet bound states. The composite excitations mix
strongly with the elementary ones in the coupling regime where 
quantum fluctuations are strong.
The quantum phase transition, known to take place in this model
at  critical frustration is
interpreted as a condensation process of (infinitely) large  
many-particle bound states.

\end{abstract}


\section{Introduction}

The $S=1/2$ quantum spin ladder is  relevant to a number of quasi
one-dimensional compounds \cite{Dagotto} and the list 
is growing as more materials become of experimental interest.
 Theoretically the two-leg ladder is, due to its geometry, the
most simple realization of a "spin-liquid" - a quantum disordered
state with gapped elementary excitations. 
The excitation spectrum of the ladder has been analyzed by a variety
of techniques, including  weak-coupling field theory mappings
 \cite{Shelton}, exact diagonalization of small clusters
\cite{Exact,Eder} and density-matrix renormalization group (DMRG)
studies \cite{DMRG}. Also,  strong-coupling
techniques have been extensively used, such as dimer series
expansions to high orders \cite{O} and mapping onto
effective bosonic theories \cite{Copalan,Eder,oleg}.
 The dispersion of the lowest triplet excitation  as well
 as the gap in the spectrum are quite well understood within the
aforementioned approaches.

 Recently, an additional branch of excitations - two-magnon
 bound  states  were found in the spin ladder model
\cite{remark,oleg}. Such bound states were also predicted
 for the dimerized quantum spin chain \cite{BS} which is another
  quantum system with a disordered ground state and gapped excitations.
    Bound states in  quasi one-dimensional gapped spin systems
 have also been observed experimentally \cite{Garrett} although
 it is still not clear which one  (or perhaps a combination
of the two above \cite{new}) is the relevant model    for their
description. 
 Two of  us have recently  pointed out  \cite{oleg} that bound
states exist, in fact, 
  in  all one and two-dimensional quantum spin systems with 
dimerization of which the spin ladder and the dimerized chain 
are particular examples.           

 In the present paper we study the two-leg spin ladder with additional
 second neighbor frustrating interactions between the chains.
  This model was introduced quite recently and analyzed
numerically via dimer series expansions \cite{us}, DMRG \cite{wang} and
exact diagonalizations \cite{us,lin}.  
 A quantum phase transition was found 
 as frustration increases from an antiferromagnetic (AF) ladder
 into  Haldane (ferromagnetic ladder) phase. The excitation spectrum
 changes dramatically as one approaches the quantum
transition
point \cite{us}.  In  a coupling region before the transition
 a singlet state appears in the triplet gap and at the transition
 both triplet and singlet gaps seem to approach zero
\cite{us,lin}.  We will show in the present work that
as frustration increases a number of low-energy many-particle 
bound states appear
in the spectrum  which mix strongly with the one-particle excitations. 
 The energies of the bound states decrease  with increasing frustration
and 
 number of particles forming them.
 Thus the quantum transition can be viewed as  softening of a very
complex excitation, composed of many-particle bound states.

Consider the Hamiltonian of  two coupled $S=1/2$ chains (spin ladder):
\begin{equation}
\label{H}
H=\sum_i\left\{J_{\perp}{\bf S}_i.{\bf S}_i^{\prime}+
J\left({ \bf S}_i.{\bf S}_{i+1}+
{\bf S}_i^{\prime}.{\bf S}_{i+1}^{\prime}\right)+
J_2\left({ \bf S}_i.{\bf S}_{i+1}^{\prime}+
{\bf S}_i^{\prime}.{\bf S}_{i+1}\right)\right\},
\end{equation}
where   
the  intra-chain ($J$) and the inter-chain ($J_{\perp}$, $J_2$) 
interactions are assumed 
antiferromagnetic $J,J_2,J_{\perp} >0$.  In Eq.(1) $J_{2}$ is
 a second neighbor inter chain coupling which causes frustration.
 In order to analyze the excitation spectrum of (1) it is convenient
to adopt the strong-coupling viewpoint. 
At large $J_{\perp}\gg J, J_{2}$ the ground state  consists
 of inter-chain  spin singlets
$|GS\rangle=|1,0\rangle |2,0\rangle |3,0\rangle ...$,
where $|i,0\rangle={1\over{\sqrt{2}}}\left[|\uparrow\rangle_i
|\downarrow\rangle_i^{\prime}-|\downarrow\rangle_i
|\uparrow\rangle_i^{\prime}\right]$. Since each  singlet can be excited into
a  triplet state it is natural to introduce a
  creation operator $t_{\alpha i}^{\dag}$  for  this 
excitation:
\begin{equation} 
|i,\alpha \rangle=t_{\alpha i}^{\dag}|i,0\rangle, \ \ \alpha=x,y,z.
\end{equation}
The representation of the spin operators in terms of $t_{\alpha i}^{\dag}$
was introduced by Sachdev and Bhatt \cite{Sachdev}: 
\begin{equation}
S_{1,2}^{\alpha} = \frac{1}{2} ( \pm   t_{\alpha} \pm
t_{\alpha}^{\dagger}   - i \epsilon_{\alpha\beta\gamma}
t_{\beta}^{\dagger}
t_{\gamma}).
\end{equation}
After 
application of this transformation to (1), or, equivalently, after 
calculating the matrix elements of the "hopping" terms $J$ and $J_{2}$ in 
(1), we find:

\begin{eqnarray}
\label{H1}
H&=&\sum_{i,\alpha,\beta}\left\{J_{\perp}t_{\alpha i}^{\dag}t_{\alpha i}+
{{\lambda}\over{2}}\left(t_{\alpha i}^{\dag}t_{\alpha i+1}+
t_{\alpha i}^{\dag}t_{\alpha i+1}^{\dag}+\mbox{h.c.}\right) \right.
\nonumber \\
& & \left. + {{\mu}\over{2}}\left(t_{\alpha i}^{\dag}
t_{\beta i+1}^{\dag}t_{\beta i}t_{\alpha i+1}
-t_{\alpha i}^{\dag}t_{\alpha i+1}^{\dag}t_{\beta i}t_{\beta i+1}\right)
\right\},
\end{eqnarray}
where we have defined

\begin{equation}
 \lambda= J-J_2, \ \   \mu=J+J_2.
\end{equation}
In addition, we have to  restrict the Hilbert space by introducing
the following  hard-core on-site 
constraint  \cite{Sachdev} 
\begin{equation}
t_{\alpha i}^{\dag}t_{\beta i}^{\dag}=0.
\end{equation}
 This exclusion of double occupancy reflects  the 
quantization of spin and ensures the uniqueness of the mapping from (1) to (4).

The Hamiltonian (\ref{H}) as well as (\ref{H1}) is symmetric under permutation
of the ladder legs. Therefore all excitations can be classified
according to this symmetry. Following standard notations we will denote
the 
antisymmetric excitations ($k_{\perp}=\pi$) by the index $u$ and
symmetric ones ($k_{\perp}=0$) by the index $g$. It is clear that the 
operator $t_{\alpha i}$ (elementary triplet) corresponds to the $u$-excitation.

The rest of the paper is organized as follows. In Section II 
we describe the one-particle (triplet) excitation spectrum.
In  Section III the two-particle problem is considered and
bound states in various channels are  analyzed.  Section IV
addresses the bound state problem for many particles focusing
mainly on the case of three particles. Section V presents 
 our analysis of the quantum phase transition in light of the previous
results and summarizes the work.

\section{Elementary triplet}
At the quadratic level the Hamiltonian (\ref{H1}) can be  diagonalized
by a combination of Fourier and   
 Bogoliubov transformations 
$ t_{\alpha k} = u_{k} \tilde{t}_{\alpha k}
+  v_{k} \tilde{t}_{\alpha -k}^{\dagger}$.
This gives the excitation spectrum:
$\omega_k^{2} = A_k^{2} - B_k^{2}$,
where $A_{k} = J_{\perp} + \lambda \cos k$ and $B_{k} = \lambda  \cos k$.
We find, in agreement with previous work \cite{Sachdev,Copalan},
that  the effect  of the quartic terms in (4) 
on the triplet spectrum is  small
and therefore we proceed by treating    these terms in mean field theory. 
 This is equivalent to
taking into account only   one-loop diagrams (first order in $\mu$).
These diagrams lead to the renormalization:
\begin{equation}
\label{AB}
A_k = J_{\perp} +(\lambda+2\mu f_1) \cos k, \ B_k = (\lambda-2\mu g_1)\cos k,
\end{equation}
where
\begin{eqnarray}
\label{fg}
f_1&=& \langle t_{\alpha i}^{\dagger}t_{\alpha i+1} \rangle =
N^{-1}\sum_q v_q^{2} \cos q \\
g_1&=& \langle t_{\alpha i} t_{\alpha i+1} \rangle =
N^{-1}\sum_q  u_q v_q \cos q .\nonumber
\end{eqnarray}
 
 The above corrections are numerically quite small.         
The dominant contribution to the spectrum renormalization 
is related to  the hard core condition Eq.(6). 
This condition is typically taken into account in the
mean-field approximation \cite{Copalan,Sachdev}.
 The latter is  
essentially uncontrolled, especially for a quasi-1D system. 
 To deal with the constraint we will use the diagrammatic approach      
 developed by us in Ref.[17].
An infinite on-site repulsion is introduced in this approach in order
to forbid the double occupancy: 
\begin{equation} 
\label{U}
H_{U} = \frac{U}{2} \sum_{i,\alpha \beta}
t_{\alpha i}^{\dagger}t_{\beta i}^{\dagger}t_{\beta i}t_{\alpha i},
 \ \ U \rightarrow \infty.
\end{equation} 
Since the interaction is infinite, the exact 
scattering amplitude  $\Gamma_{\alpha\beta,\gamma\delta}(K) =
\Gamma(K) \left(\delta_{\alpha\gamma}\delta_{\beta\delta}+
\delta_{\alpha\delta}\delta_{\beta\gamma}\right)$,
$K\equiv (k,\omega)$, 
 for the triplets has to be found.
This quantity  can be found by resumming the infinite series shown
in Fig.1(a). One can easily see that $\Gamma$ depends
on  the total energy and momentum of the incoming
particles $K = K_{1} + K_{2}$. 
The interaction (\ref{U}) is local and non-retarded which allows us
to obtain the analytic expression \cite{K,oleg}
 (in the limit  $U \rightarrow \infty$)
\begin{eqnarray}
\label{Gam}
\Gamma^{-1}(K) & = & i \int \frac{ d^{2}Q}{(2\pi)^2}
G(Q)G(K - Q) =  \nonumber \\
=&-&\frac{1}{N} \sum_q \frac{u_q^{2}
u_{k- q}^{2}}{\omega - \omega_q - \omega_{k- q}}
 + \frac{1}{N} \sum_q  \frac{v_q^{2}v_{k- q}^{2}}{\omega +
 \omega_q + \omega_{k- q}}. 
\end{eqnarray}
Here $G(Q)$ is the normal Green's function (GF)  
$G(k,t)=-i\langle T(t_{k\alpha}(t)t_{k\alpha}^{\dagger}(0))\rangle$:
\begin{equation}
\label{norm}
G(k,\omega)= \frac{u_{k}^{2}}{\omega - \omega_{k} + i\delta}
-  \frac{v_{k}^{2}}{\omega + \omega_{k} - i\delta}
\end{equation}
and the Bogoliubov coefficients  $u_k^{2}, v_k^{2} =
\pm 1/2 + A_k/2\omega_k$.
The basic approximation made  in the derivation of $\Gamma(K)$
is the neglect of all anomalous scattering vertices, which are
present in the  theory due to  the existence of anomalous GF's,
$G_{A}(k,t)=-i\langle T(t_{-k\alpha}^{\dagger}(t)t_{k\alpha}^{\dagger}(0))
\rangle$.
\begin{equation}
\label{anom}
G_{A}(k,\omega)= \frac{u_{k}v_{k}}{\omega - \omega_{k} + i\delta}
- \frac{u_{k}v_{k}}{\omega + \omega_{k} - i\delta}
\end{equation}
Our crucial observation \cite{K} is that
all   anomalous contributions are suppressed by 
a small  parameter which is present in the theory -
 the density of triplet excitations $n_{t}= \sum_{\alpha} 
\langle t_{\alpha i}^{\dagger}t_{\alpha i} \rangle = 3N^{-1}\sum_{q}
v_{q}^{2}$.         
We find that  
\mbox{$n_{t}\approx 0.1 \ (J_{\perp}/J=2)$},
 $n_{t}\approx 0.25 \ (J_{\perp}/J=1)$
and it generally increases as $J_{\perp}$ decreases.
Since  summation of ladders with anomalous  GF's
brings additional powers of $v_{q}$ into $\Gamma$, their contribution is small
compared to the dominant one of Eq.(\ref{Gam}). 
In the following analysis we will
 take into account only the contributions to the self-energy 
  which are at most linear in the triplet density  
 $n_{t}$
and therefore we  also neglect the second term in Eq.(\ref{Gam}).
 Thus our approach is expected to work as long as the gas
of triplets is dilute enough ($n_{t}$ is small).

The normal self-energy which includes  only the first power of the  
amplitude $\Gamma$ is given by the diagram in Fig.1(b):
\begin{equation}
\label{Sig}
\Sigma^{(Br)}(k,\omega) = \frac{4}{N}\sum_{q} v_q^{2}
\Gamma(k + q, \omega - \omega_q). 
\end{equation}
This is the dominant contribution to the spectrum renormalization
 as emphasized by Brueckner \cite{Mahan} who developed the 
 technique  described above  in order to study systems of
 strongly interacting fermions.

In the dilute gas approximation there are other  diagrams which
 are formally at most linear in $n_{t}$ but still numerically
give contributions much smaller than the one of Eq.(\ref{Sig}).
The first one is the ``rainbow'' correction to the anomalous self-energy
which is proportional to $\sqrt{n_{t}}$ and is shown in Fig.2(a):
\begin{equation}
\label{Sig2a}
\Sigma_A = \frac{1}{N}\sum_{q} u_q v_q
\Gamma(0,0). 
\end{equation}
This  anomalous self-energy enforces the condition 
\begin{equation}
\label{tt}
\langle t_{\alpha i}^{\dag}t_{\alpha i}^{\dagger}
\rangle = N^{-1} \sum_k u_k v_k =0.
\end{equation}
The parameters $u_k$ and $v_k$ found in the zeroth approximation
 do not satisfy
this condition. Taking into account  the self energy (\ref{Sig2a}) gives
the 
corrected values of $u_k$ and $v_k$ which do satisfy  (\ref{tt}).
 This can be seen from the formula for the renormalized Bogoliubov
coefficients, Eq.(\ref{UV}) below.
Since  (\ref{Sig2a}) is independent of $k$ and $\omega$,
technically  one can take into account the anomalous self-energy by
 introducing 
the term 
$\Lambda \sum_{i,\alpha}\left(t_{\alpha i}^{\dag}t_{\alpha i}^{\dag}+
t_{\alpha i}t_{\alpha i}\right)$ into the Hamiltonian (\ref{H1}) and
choosing the Lagrange multiplier $\Lambda$  from the condition 
(\ref{tt}).

The next correction is the contribution to the normal self-energy given
by the diagram shown in Fig.2(b), where the square denotes the scattering
amplitude (\ref{Gam}). A standard calculation gives the expression
for this diagram
\begin{equation}
\label{Sig2b}
\Sigma^{(2b)}(k,\omega) = \frac{6}{N^{2}}\sum_{p,q} 
{{(u_p v_p)(u_q v_q)u_{k+p-q}^2 \Gamma(k + p, \omega - \omega_p)
\Gamma(k + q, \omega - \omega_q)}
\over{\omega-\omega_p-\omega_q-\omega_{k+p-q}}}.
\end{equation}
Another correction is given by the diagram shown in Fig.2(c) 
plus the same diagram but with the positions of $\Gamma$ and
$\mu$ reversed. The result is 
\begin{equation}
\label{Sig2c}
\Sigma^{(2c)}(k,\omega) =-\mu \frac{4}{N^{2}}\sum_{p,q} 
{{\cos(p-q)(u_p v_p)(u_q v_q)u_{k+p-q}^2 \Gamma(k + q, \omega - \omega_q)}
\over{\omega-\omega_p-\omega_q-\omega_{k+p-q}}}.
\end{equation}
The last correction  linear in the triplet
density is shown at Fig.2(d). The corresponding expression is
\begin{eqnarray}
\label{Sig2d}
\Sigma^{(2d)}(k,\omega) =&-&\mu \frac{1}{N^{3}}\sum_{p,q,l} 
{{(u_p v_p)(u_q v_q)u_{k-p-l}^2u_{k-q-l}^2 \Gamma(k - p, \omega - \omega_p)
\Gamma(k -q, \omega - \omega_q)}
\over{(\omega-\omega_p-\omega_l-\omega_{k-p-l})
(\omega-\omega_q-\omega_l-\omega_{k-q-l})}}\\
&\times&[8\cos(p+q+l-k)+10\cos(p-q)].\nonumber
\end{eqnarray}
Let us stress again that all normal self-energy contributions
(\ref{Sig}),(\ref{Sig2b}),(\ref{Sig2c}),(\ref{Sig2d}) are quadratic in 
$v_q$ and hence linear in the triplet density. The anomalous self-energy 
(\ref{Sig2a}) is linear in $v_q$ and thus proportional to
$\sqrt{n_{t}}$.

In order to find the renormalized spectrum, one has to solve the set
of two
 coupled Dyson equations for the normal and anomalous GF's, 
shown symbolically in Fig.3. The result for the normal GF is:

\begin{equation}
G(K) = \frac{\omega + A_{k} + \Sigma(-K)}{[\omega +A_{k} + \Sigma(-K)]
[\omega -  A_{k} - \Sigma(K)] + [B_{k} + \Sigma_{A}(K)]^{2}}
\end{equation}
After separating  this equation   
 into a quasiparticle contribution and
incoherent background, we find \cite{K}:
\begin{equation}
\label{Gnn}
G(k,\omega) = \frac{Z_k U_k^{2}}{\omega - \Omega_k+i\delta} - 
\frac{Z_kV_k^{2}}{\omega + \Omega_k-i\delta} + G_{inc}. 
\end{equation}
The renormalized triplet spectrum  and the renormalization constant are:
\begin{eqnarray}
\label{OZ}
\Omega_k &=& Z_k \sqrt{[ A_k + \Sigma(k,0)]^{2}- [B_k+\Sigma_A]^{2}},\\
Z_k^{-1} &=& 1 - \left(\frac{\partial \Sigma}{\partial\omega}
\right)_{\omega =0}\nonumber .
\end{eqnarray}
Here the normal self-energy operator is given by 
Eqs.(\ref{Sig}),(\ref{Sig2b}),(\ref{Sig2c}),(\ref{Sig2d})
\begin{equation}
\label{Sigtot}
\Sigma(k,\omega)=\Sigma^{(Br)}+\Sigma^{(2b)}+\Sigma^{(2c)}+\Sigma^{(2d)}
\end{equation}
and the anomalous self-energy operator is given by Eq.(\ref{Sig2a}).
The renormalized Bogoliubov coefficients in (\ref{Gnn}) are:
\begin{equation}
\label{UV}
U_k^{2},V_k^{2} = \pm \frac{1}{2} +
\frac{ Z_k[ A_k +\Sigma(k,0)]}{2\Omega_k}.
\end{equation}
Equations (\ref{Gam},\ref{Sigtot}, \ref{OZ},\ref{UV}) have to be solved 
self-consistently for $\Sigma(k,0)$ and $Z_k$. From Eq.(\ref{Gnn}) it  
also follows  that one has to replace $u_k \rightarrow \sqrt{Z_k} U_k, \ 
v_k \rightarrow \sqrt{Z_k} V_k$ in all expressions presented above and
below.

Let us demonstrate how this approach works in the strong-coupling
limit  $J_{\perp} \gg \mu,\lambda$.
To first order in $\lambda/J_{\perp}$,  $A_k=J_{\perp}+ \lambda \cos k$ and 
$B_k=\lambda\cos k$. This leads to $\omega_k \approx A_k$,
$u_k \approx 1$, $v_k \approx - (\lambda/2J_{\perp})\cos k$
and $f_1=0$, $g_1=-\lambda/4J_{\perp}$. Substitution 
 into (\ref{Gam}), (\ref{Sig}), and (\ref{Sigtot}) gives 
\begin{eqnarray}
\label{1j}
&\Gamma(k,\omega)&=2J_{\perp}-\omega, \\
&\Sigma(k,\omega)&=\Sigma^{(Br)}(k,\omega)= 
{1\over{2}}(\lambda/J_{\perp})^{2}(3J_{\perp}-\omega).\nonumber
\end{eqnarray}
Note that self-energy corrections (\ref{Sig2a}), (\ref{Sig2b}),
(\ref{Sig2c}), and (\ref{Sig2d}) do not contribute in this order.
Then from Eq.(\ref{OZ}) we find  the quasiparticle
 residue $Z=1-(1/2)(\lambda/J_{\perp})^2$ and
the dispersion
\begin{equation}
\label{o}
\Omega_k=J_{\perp}+ \lambda \cos k +{{3 \lambda^2 }\over{4 J_{\perp}}}-
{{\lambda^2}\over{4 J_{\perp}}}\cos 2k.
\end{equation}
The result (\ref{o}) agrees with that obtained by direct $1/J_{\perp}$
expansion \cite{2} to this order. 

It is also useful to consider the  next order   in $1/J_{\perp}$.
Using  the first order calculation presented above we find
$A_k=J_{\perp}+ \lambda \cos k$ and $B_k=\lambda(1+\mu/2J_{\perp})\cos k$
and hence $u_k \approx 1$, $v_k \approx - B_k/2A_k
\approx -(\lambda/2J_{\perp})(1+\mu/2J_{\perp}) (1-\lambda/J_{\perp}
\cos k) \cos k$.
The scattering amplitude $\Gamma$ is not changed in this order and
thus given by Eq.(\ref{1j}).
The anomalous self energy calculated according to Eq.(\ref{Sig2a})
and the  contributions to the normal self-energy given by 
Eqs.(\ref{Sig}),(\ref{Sig2b}), (\ref{Sig2c}), (\ref{Sig2d}) are:
\begin{eqnarray}
\label{2nd}
&&\Sigma_A=\lambda^2/2J_{\perp}(1+\mu/2J_{\perp}),\nonumber\\
&&\Sigma^{(Br)}={1\over{2}}(\lambda/J_{\perp})^2(1+\mu/J_{\perp})
(3J_{\perp}-\omega),\nonumber\\
&&\Sigma^{(2b)}={{3\lambda^3}\over{8 J_{\perp}^2}}\cos k,\\
&&\Sigma^{(2c)}={{\mu\lambda^2}\over{4 J_{\perp}^2}},\nonumber\\
&&\Sigma^{(2d)}=-{{5\mu\lambda^2}\over{8 J_{\perp}^2}}.\nonumber
\end{eqnarray}
Substituting these into Eqs.(\ref{Sigtot}),(\ref{OZ}) we find the
elementary triplet dispersion to order $1/J_{\perp}^2$: 
\begin{equation}
\label{o1}
\Omega_k=J_{\perp}+ \lambda \cos k +{{\lambda^2 }\over{J_{\perp}}}
\left({{3}\over{4}}-{{1}\over{4}}\cos 2k\right)
+{{\lambda^3 }\over{J_{\perp}^2}}
\left(-{{1}\over{4}}\cos k+{{1}\over{8}}\cos 3k\right)
+{{\lambda^2 \mu }\over{J_{\perp}^2}}
\left({{3}\over{8}}-{{1}\over{4}}\cos 2k\right).
\end{equation}
Using Eq.(\ref{UV}) one can also prove that the condition (\ref{tt}) is
satisfied.
The result (\ref{o1}) agrees with that obtained by direct $1/J_{\perp}$
expansion \cite{O,us} to this order. 

The technique presented above   is 
certainly not the simplest way to construct  the 
$1/J_{\perp}$ expansion. Moreover it can not reproduce terms of order
 $1/J_{\perp}^3$
and higher  because  contributions to the self energies which
are  quadratic and higher order in the  triplet 
density have been neglected in our approach.
However the advantage of the method  
comes from the fact that  $n_t$ remains relatively small (0.25) even for 
$J/J_{\perp}=1$. The purpose of the presented exercise was to demonstrate 
 that the result of our approach coincides with  the result obtained
by perturbation theory around the dimer limit to the relevant order.

For arbitrary $J_{\perp}$ a self-consistent numerical solution of
 Eqs.(\ref{Gam},\ref{Sig},\ref{OZ},\ref{UV}) is required.
The triplet excitation spectra  obtained
from this solution for $J_{\perp}/J=2$ are shown in Fig.4. 
For comparison we present the spectrum for $J_{2}=0$  which only
includes
 the  Brueckner correction (\ref{Sig}) 
as well  as the spectrum which includes all terms linear in  $n_t$ 
 (the self-energies (\ref{Sig2a}),(\ref{Sig2b}),(\ref{Sig2c}),(\ref{Sig2d}),
in addition to (\ref{Sig})).
 One can see  that the Brueckner diagram is the most important one.
All other corrections are much less important,
however we will keep them in all subsequent calculations.
 Notice that the correlation corrections described above
 renormalize the spectrum
very strongly as can be seen by comparing with the bare dispersion (all
correlations neglected, $U=\mu=0$): $\omega_{k}^{2}=
J_{\perp}^{2} + 2\lambda J_{\perp} \cos k$. The bare spectrum even
becomes unstable for $J_{\perp} < 2\lambda$.
 In Fig.4  we also present for
comparison dispersions  obtained by  8-th order
 dimer series expansion \cite{O}. The agreement between our calculation and
these curves is excellent which reflects the smallness of the
triplet density $n_{t}\approx0.1$. 
 In Fig.5 we present similar plots for the case $J_{\perp}=J$.
Looking at the curves at $J_2=0$ one can say that the agreement between
our theory and the result obtained by
 series expansions is still reasonable because
the triplet density in this case is  $n_t \approx 0.25$ and hence one
has to expect about 25\% disagreement. However  as $J_2$ increases the
disagreement increases (especially at the point $k=0$) in spite of the 
fact that according to our calculation the triplet density does not increase 
and even slightly decreases. Moreover 
the excitation energy at $k=0$ vanishes at $J_2\approx 0.6J$, which
signals a quantum phase transition into the Haldane phase.
Our calculation however does not give any indication of the triplet
mode becoming soft at $k=0$. 
Therefore something important is missing in our approach.
We will demonstrate in  Section IV  that   what is missing is
the contribution of low-energy  many-particle
bound states (3,5,7... particles) which have $u$-symmetry and therefore
can mix with the elementary triplet. 

Next, we proceed with the analysis of 
two-particle bound states which  have $g$-symmetry and therefore do not mix
with the elementary triplet.

\section{Two-particle bound states}

The quartic interaction in the
Hamiltonian (\ref{H1}) leads to attraction between two triplet excitations.
We will show that the attraction is strong enough to form a singlet
(S=0) and a triplet (S=1) bound state.
The method we employ  essentially follows our previous work \cite{oleg}. 

Consider the 
scattering of two triplets:
$q_1\alpha +q_2\beta \to q_3\gamma+q_4\delta$ 
and  introduce the total ($Q$) and relative ($q$) momentum of the pair    
$q_1=Q/2+q$, $q_2=Q/2-q$, $q_3=Q/2+p$, and $q_4=Q/2-p$.
The bare (Born) scattering amplitude is (see Fig.6(a)):
\begin{eqnarray}
\label{M1}
M_{\alpha\beta,\gamma\delta}&=&
\mu \left(\delta_{\alpha\gamma}\delta_{\beta\delta}
 - \delta_{\alpha\beta}\delta_{\gamma\delta}\right)\cos(q+p)+ \nonumber \\
&& \mu \left( \delta_{\alpha\delta}\delta_{\beta\gamma} 
- \delta_{\alpha\beta}\delta_{\gamma\delta}\right)\cos(q-p) + \nonumber \\
&&U(\delta_{\alpha\gamma}\delta_{\beta\delta} + 
\delta_{\alpha\delta}\delta_{\beta\gamma}). 
\end{eqnarray}
The $\mu$ and the $U$ terms arise  from the 
quartic interaction in (\ref{H1}) and the constraint
(\ref{U}) respectively.
We also have to take into account that the triplet excitation differs
from the bare one due to the Bogoliubov transformation and the quasiparticle
residue. Therefore the following  substitution has to be made:
\begin{equation}
\label{Ms1}
M_{\alpha\beta,\gamma\delta} \to \sqrt{Z_{q_1}} U_{q_1}
      \sqrt{Z_{q_2}} U_{q_2}
      \sqrt{Z_{q_3}} U_{q_3}
      \sqrt{Z_{q_4}} U_{q_4} M_{\alpha\beta,\gamma\delta}.
\end{equation}
The  bound state satisfies the  Bethe-Salpeter equation for the  
poles of the exact scattering amplitude $\tilde{M}$. 
 This equation is presented graphically in Fig.6(b) and 
has the form \cite{remark1}:
\begin{equation}
\label{BS}
\left[E_Q-\Omega_{Q/2+q}-\Omega_{Q/2-q}\right]\psi(q)=
 \frac{1}{2}
\int {{dp}\over{2\pi}}M(Q,q,p)\psi(p).
\end{equation}
Here $M(Q,q,p)$ is the scattering amplitude in the appropriate
channel,  
 $E_Q$ is the energy of the bound state and $\psi(q)$ is the
two-particle wave
function. The factor of $2$ in Eq.(\ref{BS}) is related  to 
the symmetry of the diagram on the right hand side of Fig.6(b)
under the  exchange of the two intermediate lines. Thus in order
to avoid double counting of the intermediate states, the result has to
be divided by two. 
Let us introduce the  minimum energy for two 
excitations with given total momentum (lower edge of the two-particle
continuum) 
$E_Q^c= \mbox{min}_q \left\{\Omega_{Q/2+q}+\Omega_{Q/2-q}\right\}$.
If  a bound state exists then its energy is lower
than the continuum $E_Q < E_Q^c$. The binding energy is defined as  
$\epsilon_Q=E_Q^c-E_Q > 0$.

In the singlet (S=0) channel 
 the scattering amplitude  is:
\begin{equation}
M^{(0)}={1\over{3}}\delta_{\alpha\beta}\delta_{\gamma\delta}
M_{\alpha\beta,\gamma\delta}=-4 \mu \cos q \cos p +2U.  
\end{equation}
First,  consider the strong-coupling limit    
 $J_{\perp} \gg J,J_2 $. Let us keep  terms up to first order
in $1/J_{\perp}$, i.e. take $\Omega_{q}$ from Eq.(\ref{o}).       
The lower edge of the continuum in this order is: 

\begin{equation}
\label{ec}
E^{c}_{Q} = 2 J_{\perp} + \frac{3 \lambda^{2}}{2J_{\perp}}  +
\left\{ \begin{array}{ll}
 -\frac{\lambda^{2}}{ 2 J_{\perp}} \cos Q - 2\lambda \cos Q/2 &,
Q<Q^{*}\\
 +\frac{\lambda^{2}}{ 2 J_{\perp}}  \cos Q + J_{\perp} (\cos^{2} Q/2) /
\cos Q &, Q>Q^{*}
\end{array}
\right.
\end{equation}
Here $Q^{*}$ is determined from the equation:
$(\cos Q^{*}/2)/\cos  Q^{*} = - \lambda/ J_{\perp}$. Notice
that in the strict limit $\lambda/ J_{\perp} =0$ one has
 $Q^{*} = \pi$ and thus the upper line in Eq.(\ref{ec}) is sufficient. 
 The equation for the bound state reads:

\begin{eqnarray}
\label{BS1}
\lefteqn{\left[E_Q^{(0)}- 2 J_{\perp} -2 \lambda \cos Q/2 \cos q
 - \frac{3 \lambda^{2}}{2J_{\perp}} + 
\frac{ \lambda^{2}}{2J_{\perp}}\cos Q \cos 2q \right]\psi(q)= } \nonumber
\hspace{2cm} \\
 && =- 2\mu \cos q\int {{dp}\over{2\pi}} \cos p \psi(p)
+U\int {{dp}\over{2\pi}} \psi(p).
\end{eqnarray}
Since we work to order $1/J_{\perp}$ and 
 both $Z_{q},U_{q}= 1 + O(\lambda^{2}/J^{2}_{\perp})$, these
quantities have been set to unity in (\ref{BS1}).
 Due to the infinite repulsion ($U\rightarrow \infty$),
a Lagrange multiplier has to be introduced to enforce the
condition $\int dp \psi(p)=0$ (meaning that the bound state is
 d-wave like). The solution of Eq.(\ref{BS1}) to leading order for the
wave-function and next to leading order for the energy is:
\begin{equation}
\label{res}
\psi^{(0)}(q,Q) = \sqrt{2(1-C_Q^{2})}
\frac{\cos q +C_Q}{1+ C_{Q}^{2}+2C_Q\cos q}
+ O\left(\frac{\lambda^{2}}{\mu J_{\perp}}\right)
\end{equation}
\begin{equation}
\label{sin}
E_Q^{(0)}= 2 J_{\perp} +  \frac{3 \lambda^{2}}{2J_{\perp}}
- \mu(1+ C^{2}_{Q}) - \frac{\lambda^{2}}{4J_{\perp}} 
(1+ C^{2}_{Q}) \cos Q
\end{equation}
where we have introduced  the notation 

\begin{equation}
\label{bu}
C_{Q}={{\lambda}\over{\mu}} \cos Q/2.
\end{equation}
Thus we see that in the strong-coupling limit a singlet bound state
always exists. At $J_{\perp}=2J$, $J_2=0$  Eq.(\ref{BS}) with the 
substitution (\ref{Ms1}) has to be solved
numerically and the result is presented in Fig.7. We find that
for  $k \lesssim 2\pi/5$ the binding energy is practically zero in this
case.

In the triplet (S=1) channel   the  scattering amplitude is:

\begin{equation}
M^{(1)}={1\over{2}}\epsilon_{\rho\alpha\beta}\epsilon_{\rho\gamma\delta}
M_{\alpha\beta,\gamma\delta}=-2\mu\sin q \sin p.
\end{equation}
In this formula there is no summation over the index  $\rho$ which gives
the spin of the bound state.
By solving Eq.(\ref{BS})  in the limit $J_{\perp} \gg J,J_2$ 
we obtain for the wave-function and the binding energy:
\begin{equation}
\label{res1}
\psi^{(1)}(q,Q) =\sqrt{1/2-2C_Q^{2}}
\frac{\sin q}{1/2 + 2 C^{2}_{Q} +2C_Q\cos q} +
O\left(\frac{\lambda^{2}}{\mu J_{\perp}} \right )
\end{equation}
\begin{equation}
\label{trip}
E_Q^{(1)}=  2 J_{\perp} +  \frac{3 \lambda^{2}}{2J_{\perp}}
-\frac{\mu}{2}( 1+ 4C_Q^{2}) -  \frac{\lambda^{2}}{2J_{\perp}}
(6C_Q^{2} -1/2)\cos Q
, \ \  C_Q < 1/2. 
\end{equation}
For $C_Q >1/2$ we find that the binding energy vanishes,
 $\epsilon_Q^{(1)}= E^{c}_{Q} -E_Q^{(1)}=0$,  which means that at $J_2=0$
($\mu=\lambda=J$) the triplet bound
state only exists for momenta $k>Q_{c}=2\pi/3$ (in the strong-coupling limit)
\cite{remark}.
 At $J_{\perp}=2J$, $J_2=0$ the numerical solution of Eq.(\ref{BS}), 
plotted in Fig.7 (with the additional contribution Eq.(\ref{Lamb}))
 shows that the bound state exists down to $k\approx \pi/2$.

 Finally, we find that there is no bound state in the tensor (S=2)
channel. This is due to the fact that the  scattering
amplitude in this case 
$M^{(2)}= 2\mu \cos q \cos p + 2U$ corresponds to repulsion
 and consequently there is no solution of the Bethe-Salpeter equation
 with positive binding energy. However a solution exists with energy
above the upper edge of the two-particle continuum. In the simplest
case $J=J_{2}, \lambda = 0$ we find to leading order
 $E^{(2)} = 2J_{\perp} + \mu/2$ and thus the "anti-binding" energy
is $ \mu/2$.
 
Equation (\ref{BS}) takes into account the potential interaction
between two dressed elementary triplets, but it does not take into account
 the 
contribution of quantum fluctuations into binding. Let us consider
this effect. In the strong coupling limit the first correction to the
ground state energy of the system is due to the term 
${{\lambda}\over{2}}t_{\alpha i}^{\dag}t_{\alpha i+1}^{\dag}$ 
  in the Hamiltonian
(\ref{H1}) which virtually excites a pair of triplets. Thus the energy
correction per link to lowest order is 
\begin{equation}
\label{e0}
\delta E_0=-3{{(\lambda/2)^2}\over{2J_{\perp}}},
\end{equation}
where the coefficient 3 is due to the number of possible 
polarizations \cite{ab}.
When we have a state with a real elementary triplet, the quasiparticle 
(triplet) blocks virtual 
excitations  on two links and this increases its energy by $2|\delta E_0|$. 
 This is 
the physical origin of the third term in the dispersion (\ref{o}).
Now let us consider two quasiparticles. When they are separated by more than
one lattice spacing they block four links, but when they are  on 
nearest neighbor sites they block only three links. This gives an effective
attraction $\delta E_0$. However  two  quasiparticles  in  a
singlet (S=0) state  can virtually annihilate because  of the term 
${{\lambda}\over{2}}t_{\alpha i}t_{\alpha i+1}$  in the Hamiltonian
(\ref{H1}) which has the same tensor structure.  This term 
 gives $-\delta E_0$ and consequently
 the net  effective attraction
due to quantum fluctuations vanishes.
For the triplet (S=1) bound state there is no annihilation and therefore
the energy level shift due to blocking of quantum fluctuations 
is \cite{remark2}
\begin{equation}
\label{Lamb}
\delta E^{(1)}_Q= \delta E_0 \left|\int \sqrt{2}\sin q \ \psi^{(1)}(q,Q)
{{dq}\over{2\pi}}\right|^2.
\end{equation}
The integral  gives the probability amplitude for two quasiparticles
to be  on nearest neighbor sites. 
The two-particle triplet (S=1) bound state energy for $J_{\perp}=2J$, $J_2=0$
 is plotted in Fig.7 where the potential contribution as well as
Eq.(\ref{Lamb}) have been taken into account.
While in the strong coupling limit the binding in the triplet channel is  
weaker than the one  in the singlet channel (as can be seen from
 Eqs.(\ref{sin}),(\ref{trip})), for 
$J_{\perp}=2J$, $J_2=0$ the additional attraction due to
blocking of quantum fluctuations pushes the triplet below the singlet
 for the range of momenta $4\pi/5 \lesssim q < \pi$.

The sizes of the bound states can be determined from the
corresponding wave functions. 
As expected the size   increases with
decreasing binding energy and near the threshold  we find
$R_{rms}\sim (\epsilon)^{-1/2}, \epsilon \rightarrow 0$. 
 The self-consistent evaluation of the sizes shows that
both bound states  typically extend over a few lattice spacings
\cite{oleg}.
 
The quantity which is directly measurable in inelastic neutron scattering
experiments is the dynamical structure factor: 

\begin{equation}
 S_{g,u}(k,\omega) = 
 \int e^{i\omega t} \langle S^{g,u}_{z}(k,t)S^{g,u}_{z}
(-k,0) \rangle dt, \ \ S^{g,u}_{z,i} = S_{z,i}\pm S_{z,i}'
\end{equation}
The superscript corresponds to 
 transverse (along the rungs) momentum $k_{\perp}=0,\pi$, i.e. 
$S^{g,u}_{z,i} = S_{z,i}\pm S_{z,i}'$.
 The symmetric combination
($k_{\perp}=0$) gives the  magnetic moment of the elementary triplet
which is equal to unity. Therefore expressed in terms of Cartesian
components the magnetic moment has the form 
$M_{\mu}= - i\epsilon_{\mu\alpha\beta}t^{\dag}_{\alpha}t_{\beta}$.
This immediately gives  
\begin{equation}
\label{sym}
S_{z,i}+S_{z,i}'= -i\epsilon_{z\alpha\beta}  
t^{\dag}_{\alpha i} t_{\beta i} \to
-i\epsilon_{z\alpha\beta}\sum_q u_qv_{k-q}
\tilde{t}^{\dag}_{\alpha q} \tilde{t}^{\dag}_{\beta k-q},
\end{equation}
where we also have taken into account the  Bogoliubov transformation.
By projecting  this operator onto the bound state wave function
we find the contribution of the  $S=1$ bound state 
to the static structure factor $S_{g}(k)= \int S_{g}(k,\omega)
d \omega / 2 \pi$:  
\begin{equation}
\label{strf}
 S_{g}(k) = 4\left[\frac{1}{N}\sum_{q} \psi^{(1)}(q,k)
u_{k/2+q}v_{k/2-q} \right]^{2}= \frac{1}{2}(
\lambda/J_{\perp})^{2}\sin^{2}{k/2}(1-4C_{k}^{2}) 
+ O\left(\lambda^{4}/J_{\perp}^{4}\right). 
\end{equation}
\noindent
In this formula $C_{k}$ is defined by Eq.(\ref{bu}).
 The substitution $(u_{k},v_{k}) \rightarrow 
\sqrt{Z_{k}}(U_{k},V_{k})$ has to be made according to (\ref{Gnn})
in order to find the result for arbitrary $J_{\perp}/J$.
We have also presented the leading order of the strong coupling
expansion.

A similar calculation in the $u$ channel, i.e. for the elementary 
triplet gives  
\begin{equation}
S_{u}(k)= (u_{k}+v_{k})^{2} = 1 -\frac{\lambda}{J_{\perp}}\cos k +
 O\left(\lambda^{2}/J_{\perp}^{2}\right).
\end{equation}
 For $J_{2}=0, J_{\perp}=2J$ we have found by numerical evaluation of
the corresponding expressions that $S_{g}(\pi)/S_{u}(\pi) \approx 0.05$
and thus the experimental signal is expected to be about 20 times
weaker for the bound state compared to the elementary triplet
\cite{oleg}.

\section{Many-particle bound states}
Let us first consider a three-particle bound 
state with total spin S=1 (triplet).
This state consists of an odd number of  elementary triplets and
hence  has $u$-symmetry.
A convenient way to solve the three-particle problem is to use the variational
method. First consider the simplest ansatz: three triplet excitations
on nearest neighbor sites. Such ansatz is valid in the limit 
of zero hopping ($\lambda=0$).
A  straightforward minimization of the expectation value
of the Hamiltonian (\ref{H1}) gives the energy and the
 wave function of this state:
\begin{eqnarray}
\label{3t}
&&|k\rangle_{\rho}={1\over{\sqrt{8}}}\left(
\delta_{\alpha\rho}\delta_{\beta\gamma}+\delta_{\gamma\rho}\delta_{\alpha\beta}
\right)
\sum_n e^{ikn}
t^{\dag}_{\alpha, n-1}t^{\dag}_{\beta, n}t^{\dag}_{\gamma, n+1}
|0\rangle,\\
&&_{\rho}\langle k|H|k\rangle_{\rho}=3J_{\perp}-1.25\mu\nonumber,
\end{eqnarray}
where $k$ and $\rho$ are the momentum and the 
 polarization of the
state.
Next, one can extend this ansatz by allowing each triplet to hop  onto   
a nearby site (first order in $\lambda$):
\begin{eqnarray}
\label{3t1}
&&\psi_{\rho}(k)=a|k\rangle_{\rho}+b|k\rangle_{\rho}' ,\\
&&|k\rangle_{\rho} '={1\over{\sqrt{16}}}\left(
\delta_{\alpha\rho}\delta_{\beta\gamma}+\delta_{\gamma\rho}\delta_{\alpha\beta}
\right)
\sum_n e^{ikn}\left(
t^{\dag}_{\alpha, n-2}t^{\dag}_{\beta, n}t^{\dag}_{\gamma, n+1}+
t^{\dag}_{\alpha, n-2}t^{\dag}_{\beta, n}t^{\dag}_{\gamma, n+2}
\right)|0\rangle .
\nonumber
\end{eqnarray}
The state $\psi_{\rho}(k)$ must also be normalized, i.e. $a^{2}+b^{2}=1$.
 The Hamiltonian has to be calculated in this basis, and additionally
the energy level shifts due to blocking of quantum fluctuations
have to be included, similarly to the discussion in the previous
section. The result for the effective Hamilton matrix is:
\begin{equation}
\label{3t2}
\langle H \rangle_{eff}=
\left(
\begin{array}{cc}
3J_{\perp}-1.25\mu+2{{\lambda^2}\over{J_{\perp}}} & 
{{\lambda}\over{\sqrt{2}}}\\
{{\lambda}\over{\sqrt{2}}} & 3J_{\perp}
-\mu+{{\lambda}\over{2}}\cos k + {{17}\over{8}}{{\lambda^2}\over{J_{\perp}}}
\end{array}
\right)\nonumber
\end{equation}
Notice that the quantum fluctuation correction in the second diagonal term 
(${{17}\over{8}}{{\lambda^2}\over{J_{\perp}}}$)
is slightly larger than the one in the first term. This is the
 same effect as the one discussed in the previous section - 
effective attraction due to suppression of quantum
fluctuations. In this situation numerically this attraction is not very 
important.
The energy of the three-particle bound state is
\begin{equation}
\label{3t3}
E_3(k)=3J_{\perp}-{9\over{8}}\mu+{{\lambda}\over{4}}\cos k
+{33\over{16}}{{\lambda^2}\over{J_{\perp}}}-
\sqrt{\left({{\mu}\over{8}}+{{\lambda}\over{4}}\cos k
+{1\over{16}}{{\lambda^2}\over{J_{\perp}}}\right)^2+
{{\lambda^2}\over{2}}}.
\end{equation}
Consider first the strong coupling limit, $J_{\perp}\gg J,J_2$.
For $J_2=0$ (i.e. $\mu=\lambda=J$) eq. (\ref{3t3}) gives
\begin{eqnarray}
\label{3t4}
&&E_3(k=0)=3J_{\perp}-1.68J,\\
&&E_3(k=\pi)=3J_{\perp}-2.09J.\nonumber
\end{eqnarray}
The state with $k=\pi$ is unstable with respect to  decay into three
elementary triplets because the energy of the elementary triplet is
$\Omega_q=J_{\perp}+J\cos q$. However the state with $k=0$ is stable
with respect to this decay. Nevertheless this state is also unstable
since 
it can decay into a two-triplet bound state (Section III) and an elementary 
triplet. The threshold for this decay is $3J_{\perp}-2J$ which is
pretty close to $E_3(0)$ given by (\ref{3t4}). Therefore a quite 
natural question arises: can improvements of the variational wave
function push
the energy  $E_3(0)$ below the threshold?
To check this we extended the ansatz (\ref{3t1}) by including  states
with double hopping (order
$\lambda^{2}$): $t^{\dag}_{n-3}t^{\dag}_{n}t^{\dag}_{n+1}$,
$t^{\dag}_{n-1}t^{\dag}_{n}t^{\dag}_{n+3}$, and
$t^{\dag}_{n-2}t^{\dag}_{n}t^{\dag}_{n+2}$. We find that $E_3(0)$
decreases 
to the value $3J_{\perp}-1.77J$, but still  remains above the decay
threshold. Therefore we believe that in the strong coupling 
limit for $J_2=0$ the 
three-particle bound state does not exist. However when $J_2 > (0.3-0.4) J$ the
bound state at $k=0$ becomes stable which  follows immediately from 
 Eq.(\ref{3t3}).

For intermediate values of $J_{\perp}$ the three-particle state becomes stable
for any $J_2$. Let us consider three cases for $J_{\perp}=2J$.
According to Eq.(\ref{3t3})
\begin{eqnarray}
\label{3t5}
J_2=0: \ \ &&E_3(k=0)=5.3J,\ \
            E_3(k=\pi)=4.9J,\nonumber\\
J_2=0.4J: \ \ &&E_3(k=0)=4.4J,\ \
             E_3(k=\pi)=4.2J,\\
J_2=0.8J: \ \ &&E_3(k=0)=3.7J,\ \
              E_3(k=\pi)=3.7J.\nonumber
\end{eqnarray}
In all these cases any decay of the $k=0$ state is kinematically 
forbidden
(this can be found from comparison with the elementary triplet and two-particle
bound state spectra presented in Figs.4,7).

Next, we compare the variational results
  with numerical exact diagonalization
results we have obtained for  a $2\times10$ ladder.
Plots of the spectral function \mbox{ $A(k,\omega)=
-\pi^{-1}\mbox{Im}G(k,\omega + i\delta)$}  in the $u$-channel 
(odd number of particles) for $k=0$ found by  
Lanczos diagonalization of the Hamiltonian (\ref{H})
are presented in Fig.8. The first peak corresponds to the elementary
triplet and the second one to the three-particle bound state.
The positions of the second peak agree very well with Eq.(\ref{3t5}).
For $k=\pi$  we find numerically that a second peak is absent for
 $J_{2}=0,0.4J$ whereas  a peak with an extremely small spectral
weight seems to exist for $J_{2}=0.8J$. This can be understood from
the variational treatment since the state 
$k=\pi$ can decay into three elementary triplets (compare (\ref{3t5})
and Fig.4) for $J_2=0$. Even though this state  
is slightly below
the threshold for $J_{2}=0.4J$,
  due to the limited accuracy of our calculation
it is really hard to say whether it decays or not.
However for  $J_2=0.8J$ the state $k=\pi$ is well below the decay threshold,
and indeed a peak exists  in the corresponding spectral function.
 Thus we believe that the variational method captures quite 
accurately the main features of the spectrum.

For $J_{\perp}=J$ according to Eq.(\ref{3t3}) the three-particle bound state 
energy is
\begin{eqnarray}
\label{3t6}
J_2=0: \ \ &&E_3(k=0)=3.3J,\ \
            E_3(k=\pi)=3.0J,\nonumber\\
J_2=0.4J: \ \ &&E_3(k=0)=1.8J, \ \
             E_3(k=\pi)=1.6J,\\
J_2=0.6J: \ \ &&E_3(k=0)=1.2J, \ \
             E_3(k=\pi)=1.1J,\nonumber
\end{eqnarray}
Comparing with the exact diagonalization spectra presented in Fig.9 
and Fig.10 one can see that the 
 overall agreement is good.  Notice that while for $J_2=0$ the
variational energies are higher than the numerical ones (as one would
expect), for
$J_2=0.4J,0.6J$ they are in fact lower. We attribute  this effect to
the mixing between the three-particle and the elementary triplet
 which has not been taken into account in our approach (see the
discussion below).

In the numerical spectra  in Fig.9 and Fig.10 
  a third peak is also clearly seen.
This is the five-particle bound state. To estimate its energy as well
as the energies of bound states containing higher number of particles
we could use the $N=\infty$ approximation ($N$ is the number of
particles). In the limit $\lambda=0$ the quartic term in the
 Hamiltonian Eq.(\ref{H1}) is identical to the Hamiltonian of an
  $S=1$ Heisenberg chain with antiferromagnetic
interaction $\mu/2$. The ground state energy of the latter
(for an infinite chain) is known
quite accurately to be $-0.700742 \mu$ per link \cite{white}.
 Therefore a crude estimate for the energy of an $N$-particle
bound state (containing $N-1$ links) is
\begin{equation}
\label{5}
E_{N} = N J_{\perp} - (N-1)\times 0.7 \mu,
\end{equation}
 For the five-particle bound state 
by using the above formula and taking also into account
the increase in  energy due to blocking of quantum 
fluctuations ($3\lambda^{2}/J_{\perp}$), we obtain $E_5 \approx 4.5J$
for $J_{\perp}=J$, $J_2=0$, and $E_5 \approx 1.9J$ for
$J_{\perp}=J$, $J_2=0.4J$,
in qualitative agreement with the  numerical results presented in
 Figs.9,10.

Now we can address the problem formulated at the end of Sec.II: Why  
 the diagrammatic approach developed in
Sec.II, which works quite well for $J_{2}=0$, 
does not describe even qualitatively 
 the triplet energy spectrum for $J_{\perp}=J$
and $J_2 >0$?. 
In light of the results of the present section, we find that 
the essence of the problem is in  the neglect of
  bound states of three, five, etc. quasiparticles  
whose energies decrease with increasing  $J_2$.
 Indeed, let us fix $J_{2}=0.4J$ and 
compare the energy of the elementary triplet
  at zero momentum from Fig.5 (dashed line),
 $\Omega_{0}\approx 1.73J$, with the energies of the
 three- and five-particle bound states $E_{3}(k=0)\approx 1.8J$,
  $E_{5}(k=0)\approx 1.9J$. They are quite close, and since all 
 these states have the same quantum numbers  they mix strongly.
 Notice that  in the calculation of the one-particle properties
 as well as the three-particle problem we have not taken the
mixing into account. Thus we expect the wave function in the
$u$-sector (and similarly for the $g$-sector) to be a superposition
of states with different numbers of quasiparticles:

\begin{equation}
\label{wf}
|\Psi\rangle = Z_{1} |\Psi\rangle^{(1)} +  Z_{3} |\Psi\rangle^{(3)} 
 +   Z_{5} |\Psi\rangle^{(5)}+ \dots  
\end{equation}
In this situation the  classification of the states by the 
number of "elementary" quasi particles  is becoming 
meaningless,
and the  average number of excited triplets in the lowest excitation
at $k=0$ is increasing. 
The full description of the energy spectrum requires
 the determination of the mixing coefficients in Eq.(\ref{wf})
which is beyond the scope of the present work and will be reported
in the future.
We expect that the  energy  of the "elementary" triplet
will lower substantially at $k=0$ (with respect
to the "naive" calculation of Sec.II) due to repulsion from the
   nearby many-particle bound states. In addition,
 as can be seen from the analysis of the three- and five-particle
bound states, the larger $J_{2}$ the larger the number
of many-particle
 bound states which have low energies and mix with the "elementary"
triplet. In fact it becomes energetically more and more favorable to
form states with larger  and larger number of quasiparticles
in them as $J_{2}$ increases.
Thus we expect  that the quasiparticle residue 
  will decrease with  increasing  $J_{2}$ - an effect which
indeed can be seen from our numerical analysis (see Fig.9 for
$J_{2}=0.6J$).  Eventually a situation may occur when the
  quasiparticle residue has vanished completely which means
that very large size bound states completely dominate
in the wave function  Eq.(\ref{wf}).
  This is the
point where there is an excited triplet on every site and  
 the ground state changes its nature.

\section{Quantum phase transition in the model. 
Summary and Conclusions.}
 
 The analysis of the previous section allows us to shed new light 
onto the nature of the quantum phase transition which takes place
 in the frustrated ladder model. The phase diagram of the model 
was determined in Ref.[13] and is presented in Fig.11.
At a critical coupling $J_{2c}(J_{\perp})$ the ground state
changes from that of an antiferromagnetic (AF) spin ladder  to 
a ladder with an effective ferromagnetic interaction on the 
rungs (Haldane phase). 
 From the point of view of the triplet excitations in the AF ladder
phase, the Haldane phase is characterized by an excited 
 triplet on every rung. Thus it is not surprising that bound
states of many-particles become favorable energetically near
the quantum transition point. 

 The analysis of the energy spectrum is particularly simple
 on the line $J_{2}=J$($\lambda=0$) where  quantum fluctuations are
absent completely. It is known that  on this line 
 there is an exact eigenstate of the Hamiltonian (\ref{H1})
which is a product of singlets (dimers) on each rung \cite{exactdimer}.
This is obvious from Eq.(\ref{H1}).
 This state is the ground state in the region $J_{\perp} > 
1.4J$ (see below).  As $J_{\perp}$ decreases from a large value 
 and approaches the quantum critical point, a number of
singlet states appear in the triplet gap. Figure 12 presents 
 a plot of the elementary triplet ($u1$), two-particle singlet
($g2$), three-particle triplet ($u3$) and four-particle singlet
($g4$). The energies of these states have been  found by
analytical diagonalization of the Hamiltonian Eq.(\ref{H1}):
$E_{u1} = J_{\perp}, E_{g2} = 2J_{\perp} - 2J,
 E_{u3} = 3 J_{\perp} - 2.5J,  E_{g4} =4J_{\perp} - 4.46J$.
 It is clear that at the point $J_{2}=2J$ the two-particle singlet
crosses the one-particle triplet and thus becomes the lowest excitation
in the system. Also we observe that   the larger the number of bound
particles the larger the rate of decrease of their energy. For
comparison we have also schematically plotted the states $u9$ and $g10$.
 Thus we see that a number of singlets appear in the triplet gap
and many level crossings take place. Notice that there is no mixing
between the states since quantum fluctuations are absent ($\lambda=0$).
 At the point $J_{\perp,c}=1.4J$ the energy of the
 singlet composed of infinitely many
 quasiparticles   becomes zero, $E_{g \infty}=0$, as can be seen
from Eq.(\ref{5}).   The triplet ($u$) bound state energies do not cross 
the elementary triplet for any finite number of particles in them, 
however the  infinite particle triplet becomes degenerate with
the corresponding singlet $E_{u \infty}= E_{g \infty}=0$ at the
transition point (this also follows from  Eq.(\ref{5})).
 
 We believe that the picture of the quantum transition presented above
remains valid along the whole critical line (Fig.11).
 The transition is characterized by softening  of the singlet
and triplet (at $k=0$) modes which are basically  very large size
bound states
of  many quasiparticles in the appropriate channel.
 Slightly away from the critical line (on the AF side)
 the excitation wave function is a mixture of bound states with different
number of particles and the weight of the large-size bound states
increases as the transition is approached. 

 In summary, we have analyzed the properties of many-particle
bound states in the frustrated ladder model. We have found that
the excitation spectrum is quite complex and 
many-particle bound states
 are always present in the model. Frustration pushes the bound
states to lower energies and the effective triplet and singlet
spectra are very strongly renormalized with respect to the
simple ladder (no frustration). Thus the model is an ideal
playground for studying complex excitations in quantum spin systems.

\acknowledgements
We  would like to thank J. Oitmaa,  Z. Weihong and H.Q. Lin for
stimulating discussions  and J. Oitmaa for a critical reading
of the paper. 
 One of us  (V.N.K.) 
acknowledges financial support from the Australian Research Council.

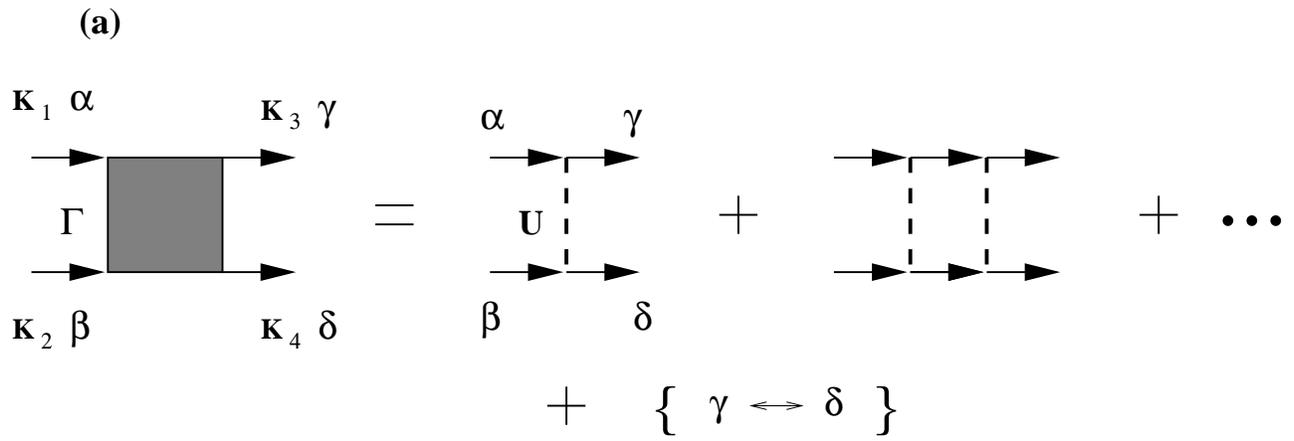
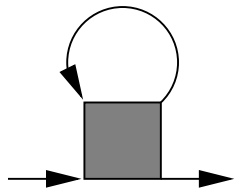
\begin{figure}
\caption
{(a) Resummation of the infinite ladder for the scattering
 amplitude $\Gamma$. The dashed line represents the (infinite)
two-particle interaction $U$.
 (b) The self-energy, corresponding to $\Gamma$.}
\label{fig.1}
\end{figure}

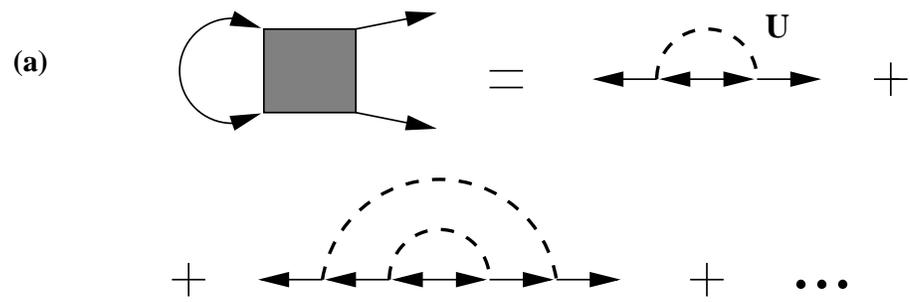
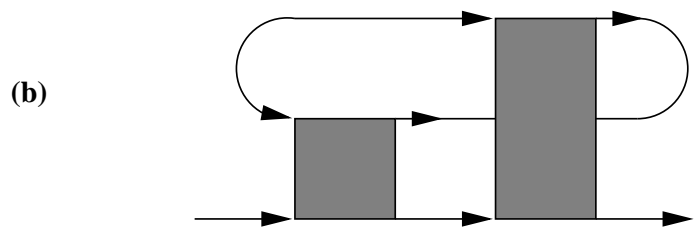
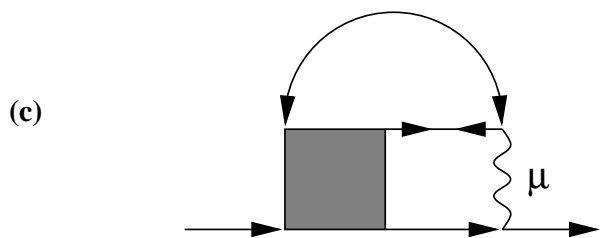
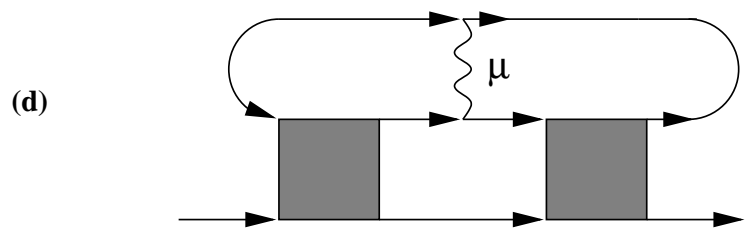
\begin{figure}
\caption
{ Diagrams for the self-energy which contribute to linear order
in the triplet density $n_{t}$. The  boxes represent the scattering
amplitude $\Gamma$ from Fig.1(a). The wavy line stands for the two-particle
interaction $\mu$, Eq.(\ref{H1}). Lines with a single arrow represent normal
Green's functions (Eq.(\ref{norm}))while lines with oppositely pointing arrows
represent anomalous Green's functions  (Eq.(\ref{anom})).}
\label{fig.2}
\end{figure}
 
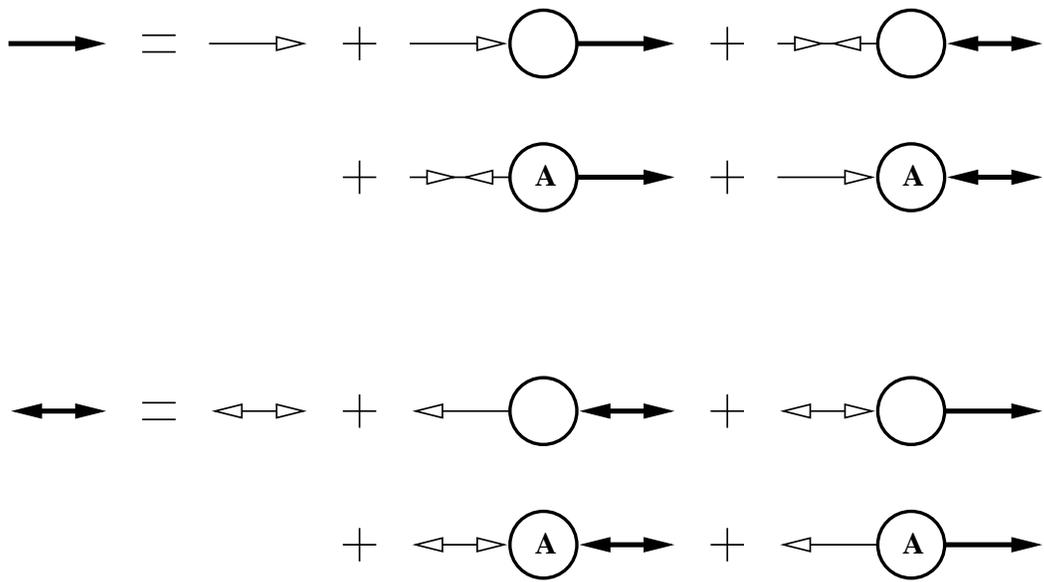
\begin{figure}
\caption
{The coupled set of Dyson's equations for the normal and anomalous
Green's functions. The anomalous self-energy (Fig.2(a)) is denoted by $A$.
The thin lines represent the bare Green's functions, Eq.(\ref{norm})
(single arrow)
and Eq.(\ref{anom}) (double arrows).}
\label{fig.3}
\end{figure}

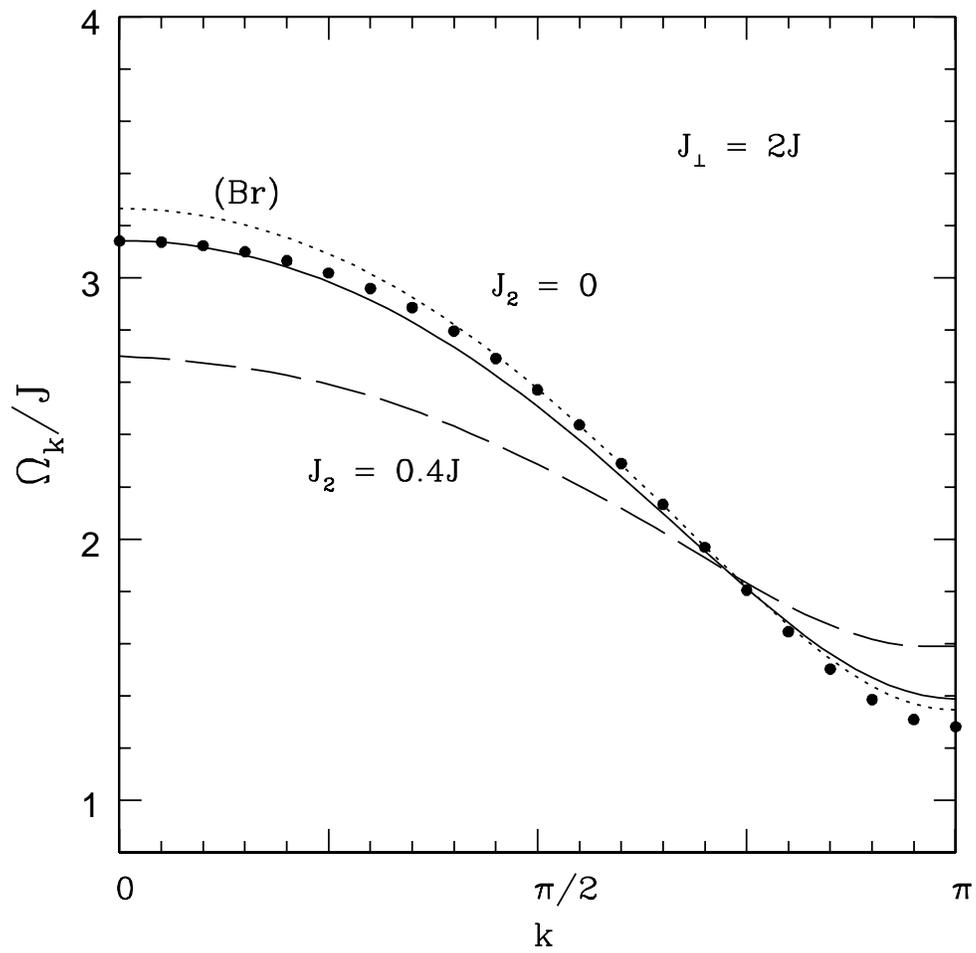
\begin{figure}
\caption
{The one-particle (triplet) excitation spectrum of the ladder 
 for $J_{\perp}=2J$. The solid dots represent numerical results
 obtained by dimer series expansions \cite{O} for $J_{2}=0$.
 The solid and dashed line are the results of the self-consistent numerical
evaluation of the spectrum Eq.(\ref{OZ})  for $J_{2}=0$ and $0.4J$,
respectively. The dotted line is the  $J_{2}=0$ result when only the Brueckner
self-energy Eq.(\ref{Sig}) is taken into account.
} 
\label{fig.4}
\end{figure}

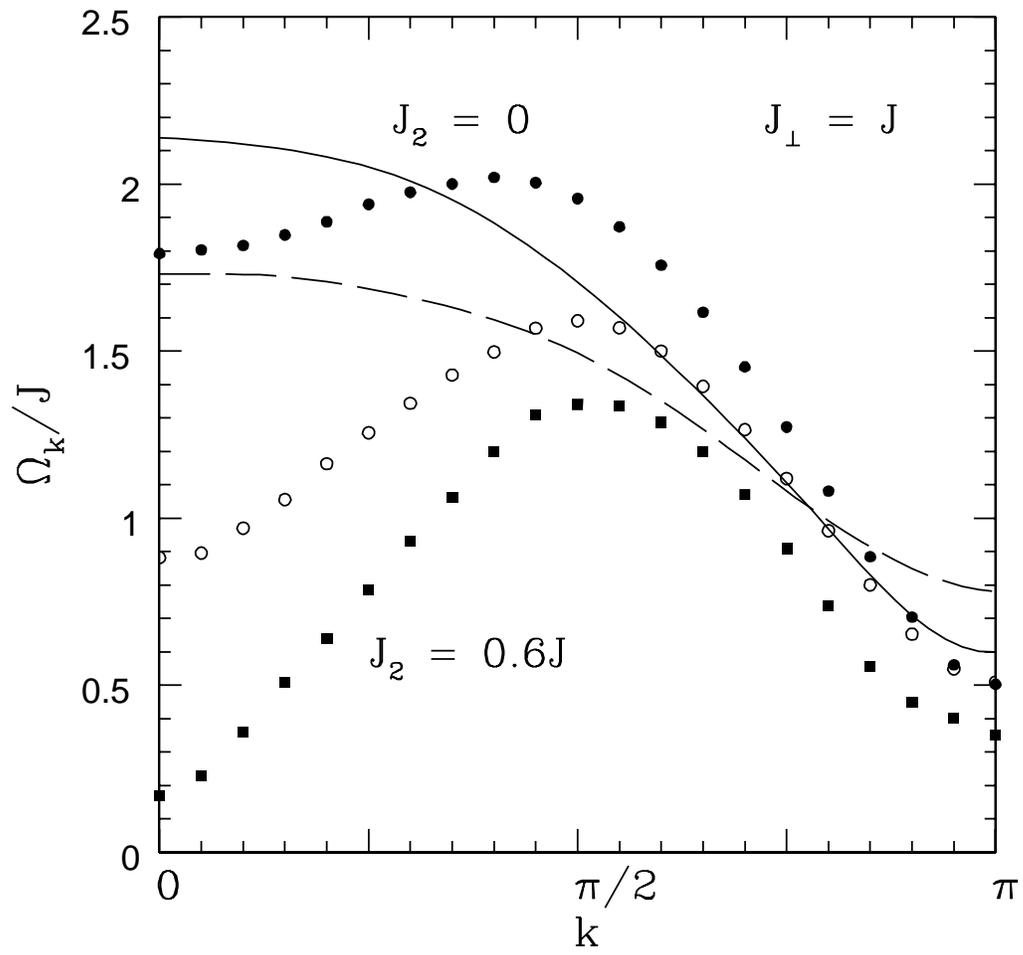
 \begin{figure}
\caption
{ One-particle spectra for $J_{\perp}=J$.
The solid dots, open circles and solid squares are
 the dimer series expansion results of Ref.[13] for
 $J_{2}=0$, $0.4J$ and $0.6J$, respectively.  The solid and dashed line
are the results of the self-consistent numerical
evaluation of the spectrum Eq.(\ref{OZ})  for $J_{2}=0$ and $0.4J$,
respectively.
}
\label{fig.5}
\end{figure}

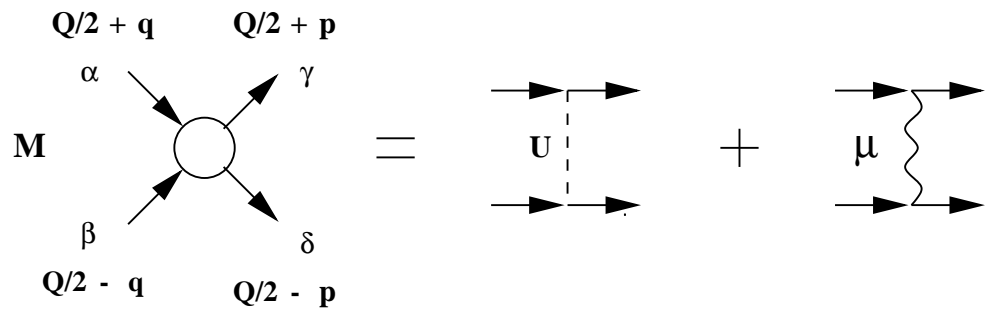
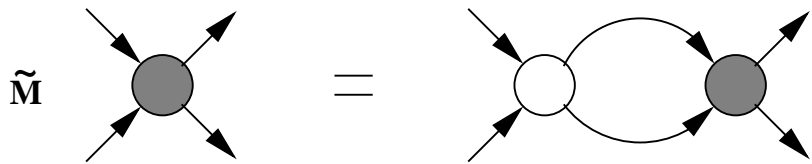
\begin{figure}
\caption
{ (a) The bare (Born) scattering amplitude $M$. 
(b) the Bethe-Salpeter equation for the poles of the exact scattering
amplitude $\tilde{M}$.
}
\label{fig.6}
\end{figure}

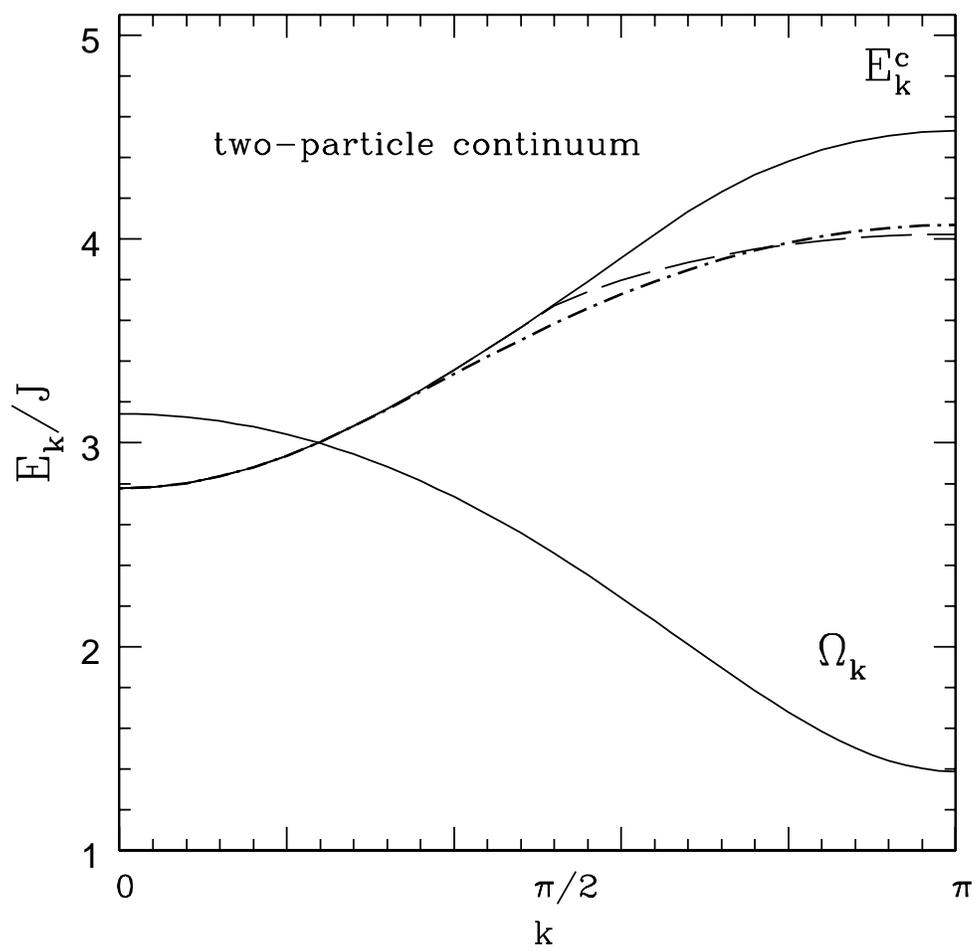
\begin{figure}
\caption
{ The excitation spectrum for $J_{\perp}=2J,J_{2}=0$ including
the  singlet  bound state (long dashed line) and the triplet bound
state (dot-dashed line). The solid line $E^{c}_{k}$ is the lower 
edge of the two-particle continuum.
}
\label{fig.7}
\end{figure}

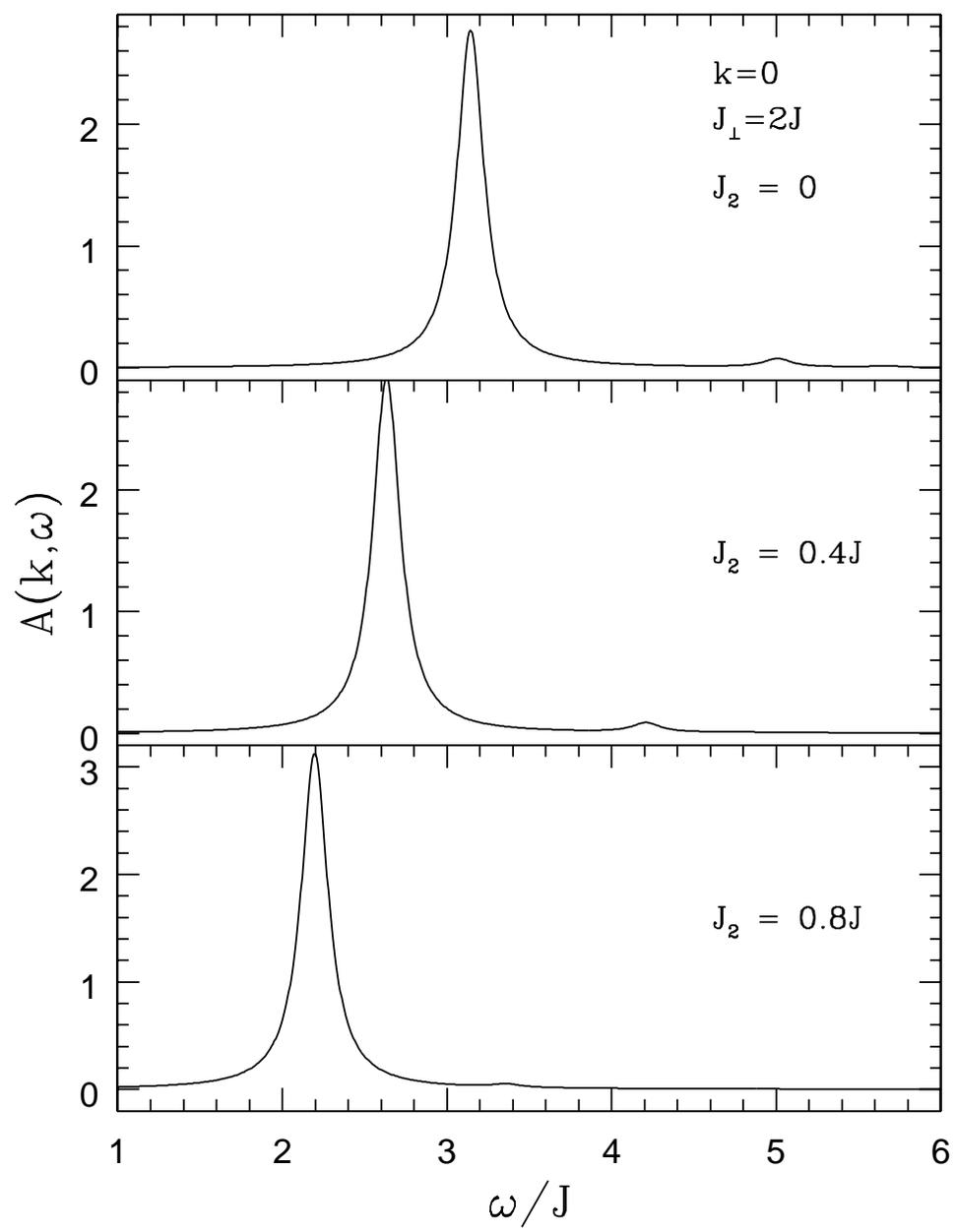
\begin{figure}
\caption
{ Spectral function $A(k,\omega)$  for $k=0,J_{\perp}=2J$ and several
values of $J_{2}$ obtained
by Lanczos diagonalization of a $2\times10$ ladder.
$\delta$-functions are replaced by Lorentzians of width $0.1J$.
}
\label{fig.8}
\end{figure}

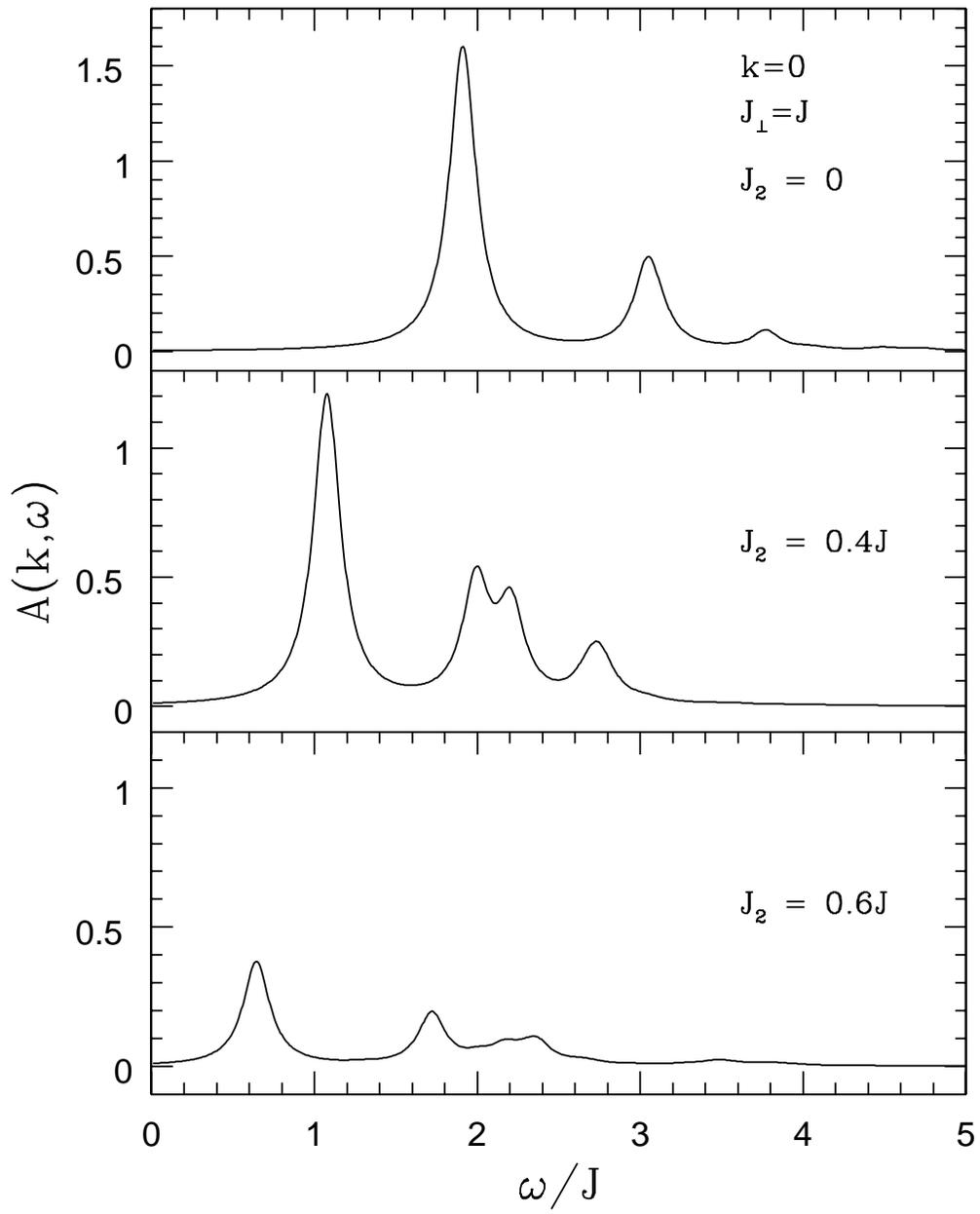
\begin{figure}
\caption
{ Same as Fig.8 for $k=0,J_{\perp}=J$.
}
\label{fig.9}
\end{figure}

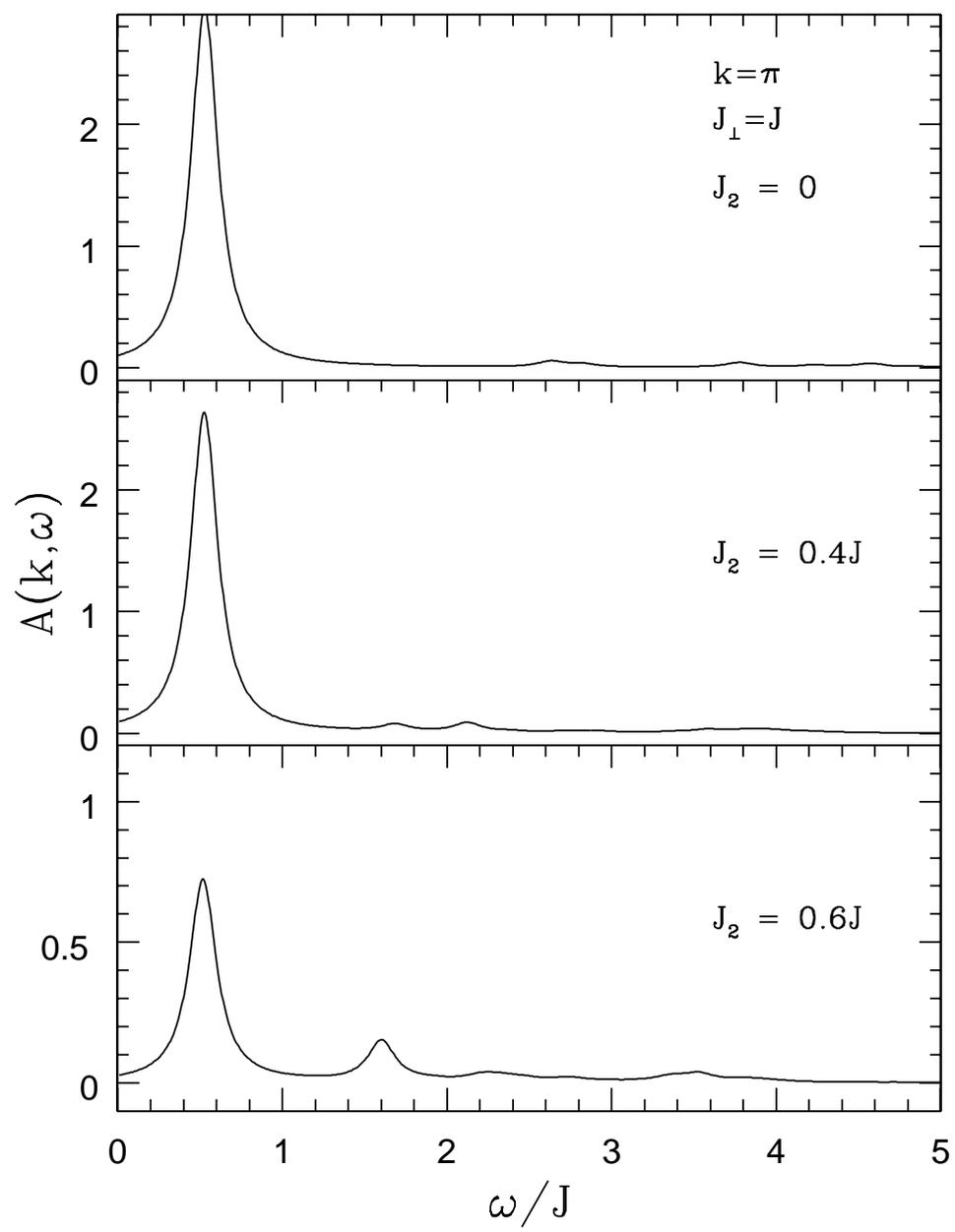
\begin{figure}
\caption
{ Same as Fig.9 for $k=\pi,J_{\perp}=J$.
}
\label{fig.10}
\end{figure}

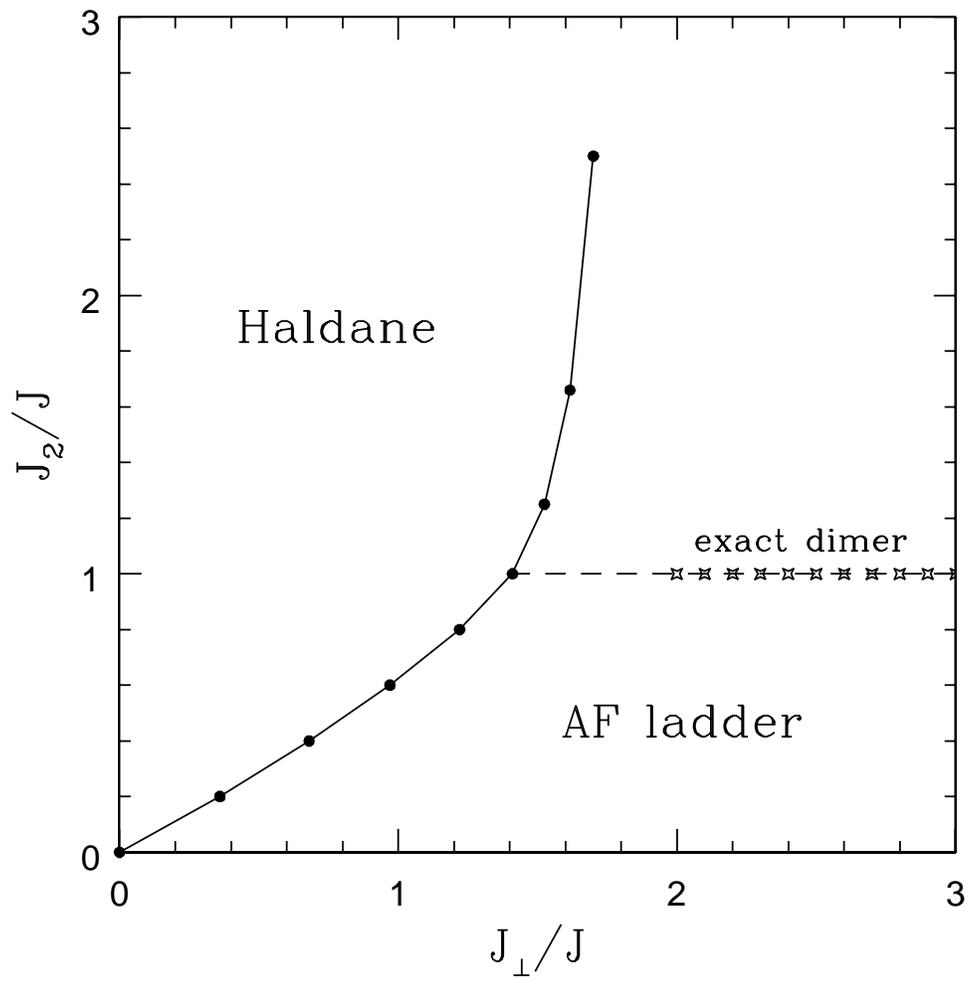
\begin{figure}
\caption
{ Phase diagram of the frustrated ladder from Ref.[13]. 
The crosses represent the line $J_{2}=J$ where the ground state is
a product of rung singlets.
}
\label{fig.11}
\end{figure}

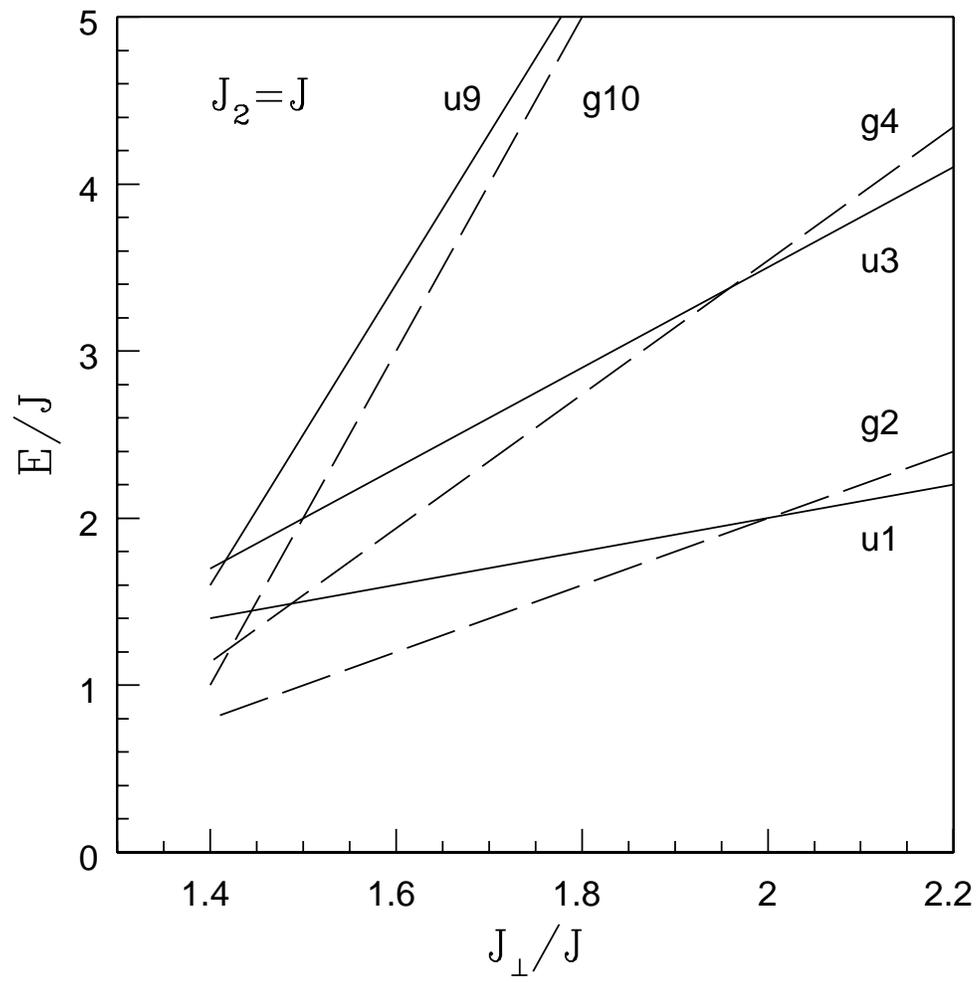
\begin{figure}
\caption
{Schematic  excitation spectrum on the line  $J_{2}=J$. 
}
\label{fig.12}
\end{figure}


\begin{references}
\bibitem{remark1}
Equation (\ref{BS}) can also be derived as a two-particle
Schr\"{o}dinger equation in momentum space $H\Psi_{Q}=E_{Q}\Psi_{Q}$.
 For the singlet bound state one has $\Psi_{Q}= \sum_{q}\psi^{(0)}
(q,Q)t_{\alpha,q_{1}}^{\dagger}t_{\alpha,q_{2}}^{\dagger}|0\rangle$,
where $q_{1,2}$ are defined in the text before Eq.(\ref{M1}).
 For  a triplet bound state with polarization $\alpha$:
$\Psi_{Q,\alpha} =  \sum_{q}\psi^{(1)}(q,Q)\epsilon_{\alpha\beta\gamma}
t_{\beta,q_{1}}^{\dagger}t_{\gamma,q_{2}}^{\dagger}|0\rangle$.
While this  way of deriving the Bethe-Salpeter equation is  quite simple and
physically transparent,  the approach presented in the text  is more general
 and we give it preference  since it is better suited for treating
more complicated interaction terms (e.g. three-body interactions
which appear in certain  models with dimerization \cite{K}).
\bibitem{ab}
Strictly speaking Eq.(\ref{e0}) is valid  for 
$J_{\perp} \gg J,J_2$. However   it works quite well 
even for $J_{\perp}\sim J$. 
For example  at the point $J_{\perp}=J$, $J_2=0$, according to 
Eq.(\ref{e0}) $\delta E_0=-0.375J$ while the numerical result\cite{O,us}  
is  $\delta E_0=-0.406J$.
\bibitem{remark2}
 This effect is similar to the shift of energy levels
 due to radiative corrections in relativistic atomic physics
 (Lamb shift).
\end{references}
\end{document}